# Electric-field independent spin-orbit coupling gap in hBN-encapsulated bilayer graphene

Fang-Ming Jing[1,2,3]†, Zhen-Xiong Shen[1,3,4]†, Guo-Quan Qin[1,2,3], Wei-Kang Zhang[1,3], Ting Lin[1,3], Ranran Cai[1,3], Zhuo-Zhi Zhang[1,2,3], Gang Cao[1,3], Lixin He[1,3,4], Xiang-Xiang Song[1,2,3]*, and Guo-Ping Guo[1,3,5]

1. CAS Key Laboratory of Quantum Information, University of Science and Technology of China, Hefei, Anhui 230026, China

2. Suzhou Institute for Advanced Research, University of Science and Technology of China, Suzhou, Jiangsu 215123, China

3. CAS Center for Excellence in Quantum Information and Quantum Physics, University of Science and Technology of China, Hefei, Anhui 230026, China

4. Institute of Artificial Intelligence, Hefei Comprehensive National Science Center, Hefei, Anhui, 230026, China

5. Origin Quantum Computing Company Limited, Hefei, Anhui 230088, China

† F.-M. J. and Z.-X. S. contributed equally to this work

* Author to whom correspondence should be addressed: X.-X. S. (songxx90@ustc.edu.cn)

# ABSTRACT

The weak spin-orbit coupling (SOC) in bilayer graphene (BLG) is essential for encoding spin qubits while bringing technical challenges for extracting the opened small SOC gap $\Delta_{SO}$ in experiments. Moreover, in addition to the intrinsic Kane-Mele term, extrinsic mechanisms also contribute to SOC in BLG, especially under experimental conditions including encapsulation of BLG with hexagonal boron nitride (hBN) and applying an external out-of-plane electric displacement field $D$. Although measurements of $\Delta_{SO}$ in hBN-encapsulated BLG have been reported, the relatively large experimental variations and existing experimental controversy make it difficult to fully understand the physical origin of $\Delta_{SO}$. Here, we report a combined experimental and theoretical study on $\Delta_{SO}$ in hBN-encapsulated BLG. We use an averaging method to extract $\Delta_{SO}$ in gate-defined single quantum dot devices. Under $D$ fields as large as 0.57-0.90 V/nm, $\Delta_{SO} \approx$ 53.4-61.8 μeV is obtained from two devices. Benchmarked with values reported at lower $D$ field regime, our results support a $D$ field-independent $\Delta_{SO}$. This behavior is confirmed by our first-principle calculations, based on which $\Delta_{SO}$ is found to be independent of $D$ field, regardless of different hBN/BLG/hBN stacking configurations. Our calculations also suggest a weak proximity effect from hBN, indicating that SOC in hBN-encapsulated BLG is dominated by the intrinsic Kane-Mele mechanism. Our results offer insightful understandings of SOC in BLG, which benefit SOC engineering and spin manipulations in BLG.

# I. INTRODUCTION

The weak spin-orbit coupling (SOC) and hyperfine interaction make bilayer graphene (BLG) promising for encoding spin qubits [1], as well as for spintronic applications [2-4]. In particular, taking advantage of the electrically generated bandgap in BLG [5,6], gate-defined quantum dots are realized [7-10], enabling manipulations of the spin degree of freedom at single-particle level. Significant efforts have been devoted to mapping spin-valley states [11-13], observing Pauli blockade [14-16], and measuring spin/valley relaxation time [17,18] in these devices, which demonstrate a promising future of BLG-based spin qubits [19].

From the view of spin manipulations, SOC is of particular importance since it has been identified as one of the key factors that leads to spin relaxation and decoherence [20,21]. In BLG, SOC is expected to be weak and is predicted to open a small gap $\Delta_{SO}$ in the low energy bands with the magnitude of tens of μeV [22,23]. Due to this small magnitude, experimental resolving $\Delta_{SO}$ is technically challenging. Recently, breakthroughs have been made so that $\Delta_{SO}$ can be experimentally extracted in both BLG and its monolayer counterpart, by means of electron-spin resonance spectroscopy [24-26] and magneto-transport spectroscopy [12,16,17,27,28]. However, the experimentally extracted $\Delta_{SO}$ not only results from the intrinsic Kane-Mele SOC mechanism in BLG, but also is affected by extrinsic mechanisms [22]. For example, BLG is usually encapsulated by hexagonal boron nitride (hBN) to maintain its pristine properties [29]. The presence of hBN and its stacking configuration with respect to the BLG layer can influence $\Delta_{SO}$ through the proximity effect [30,31]. In addition, the gate-induced out-of-plane electric displacement field $D$ can also contribute to $\Delta_{SO}$, known as the Bychkov-Rashba mechanism [22]. These extrinsic mechanisms prevent direct comparisons between measured $\Delta_{SO}$ and theoretical predictions made based on pristine BLG. Moreover, although $\Delta_{SO}$ is extracted to be in the range between 40 and 80 μeV [17,28,32,33] in hBN-encapsulated BLG, experimental controversy exists with regard to its dependency on the applied $D$ field [12,27]. It is not clear whether the different $D$ fields applied in different experiments are responsible for the observed variations in $\Delta_{SO}$. Therefore, systematic studies on $\Delta_{SO}$ in BLG under experimental conditions, including encapsulation with hBN and presence of different external $D$ fields, are highly desirable.

In this work, we report a combined experimental and theoretical study on $\Delta_{SO}$ in hBN-encapsulated BLG. First, we measure $\Delta_{SO}$ in two gate-defined single quantum dot (SQD) devices by tracing the spin-valley states of the first occupied electron under different perpendicular magnetic fields $B_\perp$. Using a developed averaging method, we obtain $\Delta_{SO} = 53.4 \pm 7.7$ μeV ($\Delta_{SO} = 60.0 \pm 13.1$ μeV) at a large $D$ field of $0.57$ V/nm ($0.63$ V/nm) in Device 1. We further extract $\Delta_{SO} = 61.8 \pm 8.4$ μeV ($\Delta_{SO} = 57.4 \pm 7.6$ μeV) at an even larger $D$ field of $0.82$ V/nm ($0.90$ V/nm) in Device 2. Benchmarked with values reported at lower $D$ fields, where ambiguous dependences of $\Delta_{SO}$ on $D$ field have been reported [12,27], our results support a $D$ field-independent $\Delta_{SO}$. Then, we perform first-principle calculations based on density functional theory (DFT) to calculate $\Delta_{SO}$ for both pristine and hBN-encapsulated BLG under external $D$ fields. We find that $\Delta_{SO}$ increases slightly when hBN encapsulation is employed, indicating a weak proximity effect. More importantly, we observe that $\Delta_{SO}$ is independent of applied $D$ field for different hBN/BLG/hBN stacking configurations, which is in good agreement with the experimental results. Our results suggest that SOC in hBN-encapsulated BLG is dominated by the intrinsic Kane-Mele mechanism, which deepens the understanding of SOC in BLG and benefits manipulating spins in BLG-based devices.

# II. METHODS AND RESULTS

Figure 1a schematically shows the band structure of hBN-encapsulated BLG when an external $D$ field is applied [12,28]. Taking the conduction band (CB) as an example, two sets of energy-degenerated but inequivalent extrema at $K$ and $K'$ points give rise to valley degree of freedom, which are connected to each other through time reversal inversion. As shown in Fig. 1a, the four lowest spin-valley states are divided into two Kramer's pairs separated by $\Delta_{SO}$. When $B_\perp$ is applied, these states split due to the spin/valley Zeeman effects with spin/valley $g$-factors of $g_s/g_v$, respectively, as shown in Fig. 1b. Therefore, tracing their evolution under $B_\perp$ yields $g_s$, $g_v$, and $\Delta_{SO}$. This can be done in an SQD device that contains only one electron, where these spin-valley states can be resolved by magneto-transport measurements. In this work, we use energy differences $\Delta E_i$ of these states to obtain $\Delta_{SO}$ (see Fig. 1c) because they are expected to be more robust against experimental fluctuations.

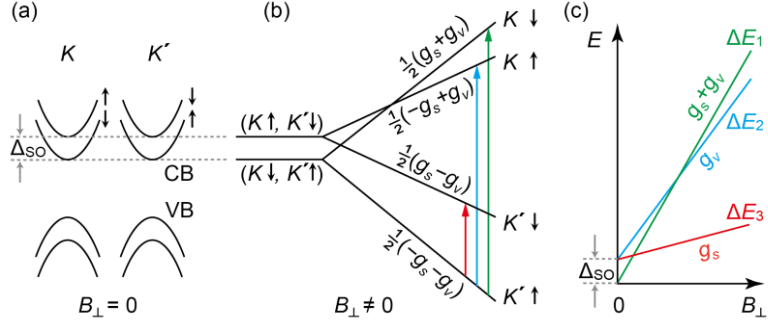

Figure1. (a) Schematic band structure of hBN-encapsulated BLG under an out-of-plane electric field. (b) When a perpendicular magnetic field $B_\perp$ is applied, the four lowest spin-valley states split due to the Zeeman effects with corresponding $g$-factors. (c) $\Delta_{SO}$ can be extracted by fitting $\Delta E_i$ as a function of $B_\perp$. Here $\Delta E_i$ denote the energy differences between the ground state ($K'\uparrow$) and three excited states, labeled by corresponding colored arrows in (b).

Figure 2a shows the cross-sectional schematic of the SQD device used for extracting $\Delta_{SO}$. Source and drain electrodes are edge-contacted to hBN-encapsulated BLG. Opposite voltages are applied to the back and split gates to generate an out-of-plane $D$ field to make BLG underneath the two split gates insulating. The finger gates are responsible for manipulating the potential landscape in the conducting channel left in between. More information about the fabrication process can be found in section S1, Supplemental Material [34].

First, taking Device 1 as an example, we define an electron SQD underneath the second finger gate (labeled as FG) using natural p-n junctions as tunnel barriers [43]. Here we apply $V_{BG} = -5.5$ V, $V_{SG1} = 4.05$ V, and $V_{SG2} = 4.02$ V to generate a large $D$ field of $0.57$ V/nm (see section S2, Supplemental Material [34] for $D$ field estimation). Figure 2b shows the first two Coulomb diamonds (see section S3, Supplemental Material [34] for more details on addressing the electron occupation number), where the addition energy $E_{add} = 7.9$ meV and the lever arm $\alpha = 0.025$ eV/V are obtained. Device 1 is characterized at a temperature of $\approx 20$ mK.

Next, we focus on the first occupied electron to investigate its spin-valley states upon varying $B_\perp$. Figure 2c shows a zoom-in of the region near the first Coulomb resonance peak in the absence of $B_\perp$. A series of parallel transition lines are observed (see colored arrows), corresponding to loading the first electron to its ground and excited states, respectively. At $B_\perp = 0$ T, the separation between the four lowest spin-valley states is $\Delta_{SO}$, with a typical value of tens of μeV. Therefore, they are expected to locate closely in the Coulomb diamond diagram (indicated by the black arrow). The transition line indicated by the yellow arrow is attributed to tunneling via higher orbital excited states, from which a typical orbital energy $\Delta_{\text{orbit}} \approx 1.6$ meV is obtained. It provides an estimation of the dot diameter $d \approx 54$ nm according to $d = \sqrt{2\hbar^2/(m^*\Delta_{\text{orbit}})}$ (Ref. [8]), where $m^* = 0.033\,m_e$ is the effective electron mass in BLG [44]. This value is in good agreement with our device design (see geometric parameters in section S1, Supplemental Material [34]). When non-zero $B_\perp$ is applied (for example $B_\perp = 0.54$ T in Fig. 2d), the spin-valley states are shifted by $B_\perp$ due to the spin and valley Zeeman effects. Since $g_v$ is typically one order of magnitude larger than $g_s$ [45], the four lowest spin-valley states evolve into two valley dominant branches. This results in two sets of transition lines, as indicated by black and magenta arrows in Fig. 2d. Increasing $B_\perp$ to 0.99 T leads to a larger separation between these two branches (see Fig. 2e). Fitting the evolution of these resonance peaks as a function of $B_\perp$ yields $g_s$, $g_v$, and $\Delta_{\text{SO}}$. Here we use $\Delta V_b$ and $\Delta V_{\text{FG}}$ for plotting Figs. 2c-e, by subtracting voltage drifts accumulated during obtaining each figure. We would like to point out that slight enhancements of conductance appear parallel to the resonance peaks (see, for example, the purple triangle in Fig. 2c). However, their distances with respect to the ground resonance peak remain constant upon varying $B_\perp$, thus their influence on tracing the evolution of the resonance peaks is limited. We attribute these unexpected conductance enhancements to an unintentional quantum dot trapped nearby, which is weakly coupled to the investigated dot and the corresponding gate electrode [46].

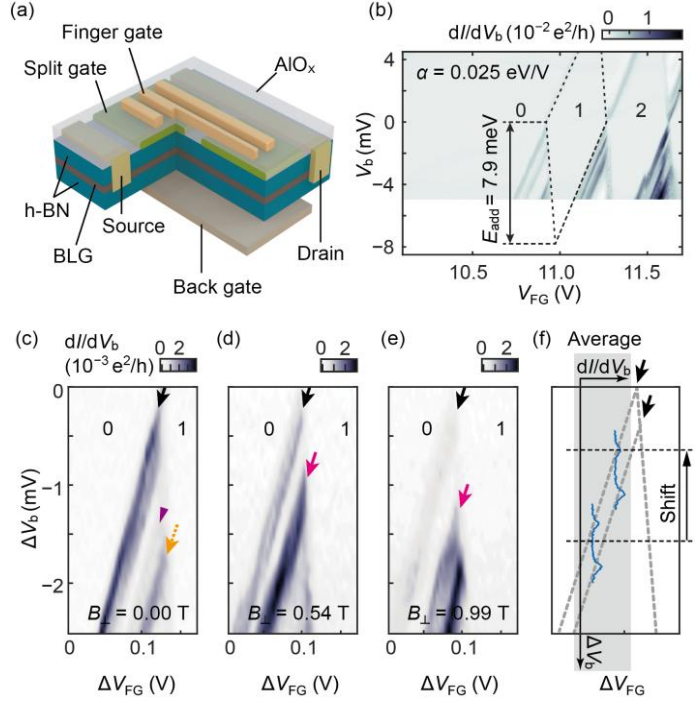

Figure 2. (a) Cross-sectional schematic of the SQD device based on hBN-encapsulated BLG. (b) Coulomb diamonds measured from Device 1 at $V_{\text{BG}} = -5.5$ V, $V_{\text{SG1}} = 4.05$ V, and $V_{\text{SG2}} = 4.02$ V. (c-e) Zoom-in near the first Coulomb resonance peak, highlighting transition lines of the first electron at (c) $B_\perp = 0.00$ T, (d) $B_\perp = 0.54$ T, and (e) $B_\perp = 0.99$ T, respectively. Black and magenta arrows indicate parallel transition lines corresponding to the four lowest spin-valley states, while the yellow arrow in (c) indicates the higher orbital excited state. The purple triangle in (c) marks the slight conductance enhancement which may originate from an unintentional quantum dot trapped nearby. (f) Schematic illustration of the averaging method. Differential conductance curves measured at different $\Delta V_{\text{FG}}$ are shifted to make transition peaks aligned and then averaged to increase the signal-to-noise ratio.

In order to better extract $\Delta_{SO}$ from fitting, we develop a method to average the data to increase the signal-to-noise ratio. As illustrated in Fig. 2f, since the transition lines are parallel to each other (two gray dashed lines highlighted by black arrows), the distance between transition peaks in differential conductance curves (two blue curves) remains constant while varying $\Delta V_{FG}$. Therefore, we are able to shift the differential conductance curves measured at different $\Delta V_{FG}$ to make transition peaks aligned. Then we average the data of different curves to suppress background fluctuations thus increasing the signal-to-noise ratio. After that, peak distances are extracted from the averaged data and are converted to energy differences $\Delta E_i$ accordingly. Finally, we repeat the procedure for different $B_\perp$ and fit $\Delta E_i$ as a function of $B_\perp$ to obtain $\Delta_{SO}$. More information about the averaging method can be found in section S4, Supplemental Material [34].

Figure 3a shows extracted energy differences $\Delta E_1$, $\Delta E_2$ and $\Delta E_3$ at different $B_\perp$. Although we have employed the averaging method, we are not able to resolve all resonance peaks for each $B_\perp$ field due to the small energy separations, compared with the line widths of the resonance peaks (see section S5, Supplemental Material [34]). Here we only use the data points when peaks can be clearly identified for extracting $\Delta E_i$. Therefore, they tend to concentrate at larger $B_\perp$ since peak separations increase with $B_\perp$. The slopes of $\Delta E_i$ correspond to spin and valley $g$-factors, respectively (see Fig. 1c). For example, $g_v = 17.6 \pm 0.9$ and $g_s = 2.0 \pm 0.2$ are obtained from fitting $\Delta E_2$ and $\Delta E_3$ as a function of $B_\perp$, respectively. The values are consistent with $g_v + g_s = 19.7 \pm 0.9$ by fitting $\Delta E_1$ and are in good agreement with previously reported $g_s$ [7,47] and $g_v$ [12,48], respectively. More importantly, the intercept of $\Delta E_3$ corresponds to $\Delta_{SO}$. Therefore, by fitting $\Delta E_3$, we extract $\Delta_{SO} = 53.4 \pm 7.7$ μeV. We would like to point out that fitting $\Delta E_2$ also yields $\Delta_{SO}$, but with a larger uncertainty ($71.7 \pm 35.8$ μeV). This may originate from the crossover of $\Delta E_1$ and $\Delta E_2$ at the intermediate magnetic field regime. In this regime, energies of $K\downarrow$ and $K\uparrow$ states are close, thus only three peaks can be resolved (see Fig. 3a).

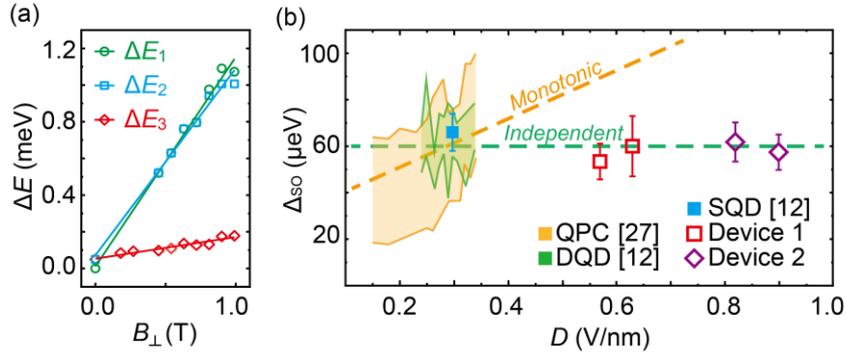

Figure 3. (a) Extracted energy differences $\Delta E_1$, $\Delta E_2$ and $\Delta E_3$ as a function of $B_\perp$ at $D = 0.57$ V/nm. Solid lines are linear fits to the data points, from which $\Delta_{SO} = 53.4 \pm 7.7$ μeV is obtained. (b) Benchmarking of our extracted $\Delta_{SO}$ (red open squares and purple open diamonds for Device 1 and Device 2, respectively) with those measured from quantum point contact (QPC, yellow shaded region), double quantum dot (DQD, green shaded region), and single quantum dot (SQD, blue square) devices. Our data points at larger $D$ fields support a $D$ field-independent $\Delta_{SO}$ in hBN-encapsulated BLG. Dashed lines are guides-to-the-eye.

It has been predicted that in addition to the intrinsic Kane-Mele SOC coupling [22,49], which is expected to be independent of $D$ field, extrinsic mechanisms also contribute to SOC under experimental conditions. These extrinsic mechanisms originate from, for example, the presence of experimental substrates and external $D$ fields, which can behave differently upon increasing $D$ field [22,31]. Previous reports suggest $\Delta_{SO}$ in hBN-encapsulated BLG ranges from 40 to 80 μeV [12,17,27,28,33]. These values are measured from different devices where $D$ fields can be varied. Moreover, at lower $D$ field regime (0.15-0.34 V/nm), a monotonic dependence of $\Delta_{SO}$ on $D$ field is reported in a quantum point contact (QPC) device [27], while an independent behavior on $D$ field is observed in double quantum dot (DQD) devices [12]. Due to the small magnitude of $\Delta_{SO}$, experimental variations make it challenging to clearly resolve the behavior of $\Delta_{SO}$ when varying $D$ field, especially at lower $D$ field regime. Therefore, we benchmark our results obtained at large $D$ fields, where the difference between monotonic and independent behaviors becomes pronounced, with those values obtained at lower $D$ fields (see Fig. 3b). In addition to $\Delta_{SO} = 53.4 \pm 7.7$ μeV extracted at $D = 0.57$ V/nm, we also obtain $\Delta_{SO} = 60.0 \pm 13.1$ μeV at $D = 0.63$ V/nm from Device 1 (see section S6, Supplemental Material [34]). Under even larger displacement fields, we obtain $\Delta_{SO} = 61.8 \pm 8.4$ μeV at $D = 0.82$ V/nm, and $\Delta_{SO} = 57.4 \pm 7.6$ μeV at $D = 0.90$ V/nm, respectively, in another device (Device 2) (see section S7, Supplemental Material [34]) at a temperature of ≈270 mK. Our results at large $D$ fields clearly support a $D$ field-independent $\Delta_{SO}$.

We would like to point out that although more experiments on estimating $\Delta_{SO}$ in the range of 40 to 80 μeV have been reported [13-15,17,18,28,32,33], exact values of the $D$ field are not provided. These values are difficult for us to estimate, since they are strongly sensitive to the thickness of hBN flakes, voltages applied to gate electrodes, and especially, gate voltage offsets due to unintentional doping (see section S2, Supplemental Material [34]). Therefore, we only use data points which their corresponding $D$ field can be extracted for benchmarking. These results are obtained from devices with similar architectures. It is also worth noting that although Device 1 and Device 2 are measured at different base temperatures, their difference in effective electron temperature is much smaller (both in the same order of 100 mK in our systems), thereby the temperature-induced influence is not obvious.

To better understand the physics of $\Delta_{SO}$ in BLG, we perform first-principle calculations based on DFT implemented in the Atomic Orbital Based Ab-initio Computation at UStc (ABACUS) [50,51]. The band structures of both pristine Bernal-stacked BLG (see Fig. 4a) and hBN-encapsulated BLG (see Fig. 4b) are calculated under different $D$ fields, from which $\Delta_{SO}$ is extracted (more information about the calculations can be found in section S8, Supplemental Material [34]). Figure 4c shows the calculated $\Delta_{SO}$ when varying $D$ field for pristine and hBN-encapsulated BLG, respectively. Two main features are found. First, $\Delta_{SO}$ in hBN-encapsulated BLG is slightly larger than that in pristine BLG, indicating a weak proximity effect from hBN. Second, $\Delta_{SO}$ is independent of $D$ field for both the conduction band (CB) and the valence band (VB), and for both pristine and hBN-encapsulated BLG, which is in good agreement with the benchmarked results shown in Fig. 3b. These two features suggest that the intrinsic Kane-Mele mechanism dominates $\Delta_{SO}$ in BLG, even when hBN encapsulation and external $D$ fields are employed.

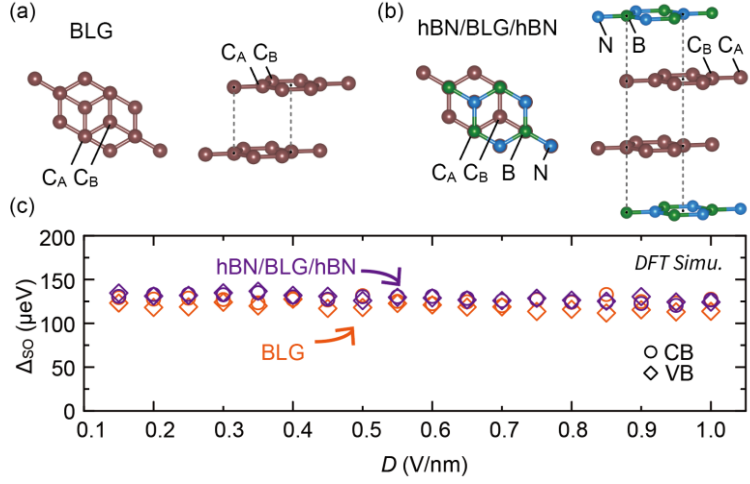

Figure 4. Stacking configurations of (a) pristine BLG and (b) hBN-encapsulated BLG for first-principle calculations. (c) Calculated $\Delta_{SO}$ as a function of $D$ field, showing a $D$ field-independent $\Delta_{SO}$ for both pristine and hBN-encapsulated BLG. CB and VB correspond to the conduction band and the valence band, respectively.

### III.  DISCUSSIONS

There are several points that we would like to clarify and emphasize in our studies.

1. It is worth noting that the calculated $\Delta_{SO}$ in Fig. 4c is larger than the measured $\Delta_{SO}$ benchmarked in Fig. 3b. This can be explained since the specific choice of pseudopotentials and structural relaxation can influence the obtained band structure [52] thus $\Delta_{SO}$, especially for the small energy scale of few tens of μeV. In our calculations, $\Delta_{SO}$ can be rescaled by adjusting a normalized SOC constant $\eta_{SO}$ embedded in the model (see section S9, Supplemental Material [34]). We would like to emphasize that the choice of $\eta_{SO}$ does not affect the result of a $D$ field-independent $\Delta_{SO}$. Therefore, we keep $\eta_{SO} = 1$ in the calculations for simplicity.

2. When applying a $D$ field to BLG, the trigonal warping effect evolves [53-55]. The conduction band minimum (CBM) and valence band maximum (VBM) are no longer located at $K/K'$ points. In our calculations, we take this influence into consideration and find $\Delta_{SO}$ remains almost unchanged from $K/K'$ points to CBM (VBM) for both pristine and hBN-encapsulated BLG when varying external $D$ fields (see section S10, Supplemental Material [34]). Thus, the influence of the trigonal warping effect on $\Delta_{SO}$ can be neglected and we use the data points corresponding to CBM and VBM for plotting Fig. 4c.

3. When benchmarking results obtained from different devices in Fig. 3b, there may be the possibility that the $\Delta_{SO}$-$D$ field relation varies from device to device since the stacking configurations of hBN with respect to BLG are randomly selected during encapsulation in experiments. As a consequence, the influence from the adjacent hBN layers can vary in different devices, resulting in different $\Delta_{SO}$-$D$ field relations. To address this possibility, we calculate $\Delta_{SO}$ as a function of $D$ field for five additional typical hBN/BLG/hBN stacking configurations with high symmetry (see section S11, Supplemental Material [34]) and find that this hypothesis is not likely to be true. Indeed, the exact values of $\Delta_{SO}$ can vary for different stacking configurations, probably due to variations in sublattice symmetry breaking [25,26] when stacking hBN and BLG in different manners. However, the calculated $\Delta_{SO}$ is found to be independent of $D$ field, regardless of stacking configurations. In a real experimental device, where twist angles between BLG and hBN layers can play a role, the overall measured $\Delta_{SO}$ is expected to result from the average of local values of the different local stacking configurations [30,31], thus should be independent of $D$ field as well.

4. According to our calculations, $\Delta_{SO}$ can be different between CB and VB for particular stacking configurations (see section S11, Supplemental Material [34]). However, this difference is expected to be suppressed in real devices if considering the influences of rescaling by $\eta_{SO}$ and averaging among different stacking configurations. This may explain the observed particle-hole symmetric $\Delta_{SO}$ [33].

5. We would like to point out that our work invites further investigations. Previous studies reveal SOC induced zero-field splitting increases when lowering the temperature from 12 K to 2 K in hBN-encapsulated BLG [26]. This behavior is attributed to local strain accumulation when depositing metal electrodes onto hBN/BLG/hBN and electron-phonon coupling in the heterostructure, which are clearly relevant in our experiments while are absent in our first-principle calculations. In addition, it is beneficial to directly include the twist angle between BLG and hBN in calculations [31]. From the experimental point of view, $\Delta_{SO}$ has been measured to be $42.2 \pm 0.8\ \mu eV$ in monolayer graphene on trenched $SiO_2$ [24]. Similarly, we can expect measuring $\Delta_{SO}$ in BLG using suspending SQD devices [56,57] to avoid the influence of the substrate. Alternatively, studies based on BLG with controlled twist angle respect to the hBN substrate [58,59] and BLG supported by various substrates [60,61] can also be expected to engineer SOC in BLG. Finally, the method developed in this work can be applied to investigate SOC of other two-dimensional materials, such as transition metal dichalcogenides, based on which quantum dots have been extensively studied [62-67].

IV. **SUMMARY**

In summary, we measure $\Delta_{SO}$ in hBN-encapsulated BLG using SQD devices. Employing an averaging method, we extract $\Delta_{SO} = 53.4 \pm 7.7\ \mu eV$ at $D = 0.57\ V/nm$ and $\Delta_{SO} = 60.0 \pm 13.1\ \mu eV$ at $D = 0.63\ V/nm$ from Device 1, and $\Delta_{SO} = 61.8 \pm 8.4\ \mu eV$ at $D = 0.82\ V/nm$ and $\Delta_{SO} = 57.4 \pm 7.6\ \mu eV$ at $D = 0.90\ V/nm$ from Device 2. Benchmarked with previous results obtained at lower $D$ fields, our results support a $D$ field-independent $\Delta_{SO}$. To confirm this behavior, we perform first-principle calculations on $\Delta_{SO}$ under different external $D$ fields. We find that $\Delta_{SO}$ is independent of $D$ field, regardless of hBN/BLG/hBN stacking configurations, which is in good agreement of the experimental results. In addition, the proximity effect is found to be weak according to the calculations, indicating that the intrinsic Kane-Mele mechanism dominates SOC in hBN-encapsulated BLG. Our work provides insightful understandings of SOC in BLG, which are essential for manipulating spins to encode qubits based on BLG.

**Note:** During the submission of our work, we became aware of a separate work [68], in which $\Delta_{SO}$ is extracted to be $59\ \mu eV$ at $D = 0.9\ V/nm$ using a different time-resolved charge detection technique, while the $D$ field dependency is not discussed.

**Data availability:** All data generated during this study are included in this article and its Supplemental Material or are available from the corresponding author upon reasonable request.

# ACKNOWLEDGMENTS

This work was supported by the National Natural Science Foundation of China (Grant Nos. 12274397, 11904351, 12274401, 92265113, 12134012, 12304560, 12404130 and 12034018), the Natural Science Foundation of Jiangsu Province (Grant No. BK20240123), and China Postdoctoral Science Foundation (Grant Nos. BX20220281 and 2023M733408). This work was partially carried out at the USTC Center for Micro and Nanoscale Research and Fabrication. The numerical calculations were performed on the USTC HPC facilities and Hefei Advanced Computing Center.